\documentclass[prl,twocolumn,amssymb,showpacs,floatfix]{revtex4}
\usepackage{graphicx}
\newcommand{\bea}{\begin{eqnarray}}
\newcommand{\eea}{\end{eqnarray}}
\newcommand{\be}{\begin{equation}}
\newcommand{\ee}{\end{equation}}
\newcommand{\rmscr}[1]{{\hbox{\scriptsize \rm{#1}}}}
\newcommand{\rmmat}[1]{{\hbox{\rm{#1}}}}
\def\apj{Astrophys. J.}
\def\apjl{Astrophys. J. Lett.}
\def\apjs{Astrophys. J. Suppl. Ser.}
\def\apss{Astrophys. Space Sci.}
\def\araa{Annu. Rev. Astron. Astrophys.}
\def\physrep{Phys. Rep. }
\def\ie{{\em i.e.\ }}
\def\eg{{\em e.g.\ }}
\def\etal{{\em et al.\ }}

\def\rmmat#1{{\hbox{\rm #1}}}
\def\rmscr#1{\rmmat{\scriptsize #1}}
\begin{document}

\title{Electron Positron Capture Rates and $\beta$-Equilibrium Condition 
for Electron-Positron Plasma with Nucleons}
\author{Yefei Yuan} 
\affiliation{Center for Astrophysics, University of Science and Technology
	of China, Hefei, Anhui 230026, P.R. China}
\affiliation{Harvard-Smithsonian Center for Astrophysics, MS-51, 60
Garden Street, Cambridge 02138}

\begin{abstract}
The exact reaction rates of beta processes for all particles 
at arbitrary degeneracy are derived, and
an {\it analytic} $\beta$-equilibrium condition $\mu_n=\mu_p+2\mu_e$
for the hot electron-positron plasma with nucleons is found,
if the matter is transparent to neutrinos.
This simple analytic formula is valid only if electrons are
nondegenerate, which is generally satisfied in the hot
electron-positron plasma. Therefore, it can be used to efficiently 
determine the steady state of the hot matter with positrons.

\end{abstract}

\maketitle

Gamma Ray Bursts (GRBs) \citep{2002ARA&A..40..137M} and 
core collapse supernovae(SNe) \citep{2001pnsi.conf..333J}
are two of the most violent events in our universe. Ironically,
their explosion mechanisms are still mysterious. 
Recently, the central engine of GRBs is believed to be related to 
the hyperaccretion of a stellar-mass black hole at extremely high rates
from $\sim$ 0.01 to 10 $M_{\odot}$s$^{-1}$ 
\citep{1992ApJ...395L..83N,2003ApJ...586.1254P,2003ApJ...588..931B}.
In such an accretion disk,
matter is so dense that photons are trapped. The possible 
channel for energy release is either neutrino emission 
which is mainly from the electron-positron ($e^{\pm}$) capture on nucleons and 
$e^{\pm}$ annihilation, or outflows from the disk.
Whatever a successful central engine is, it ejects a hot
fireball which consists of the radiation field 
and baryons. 
The ratio of neutrons to protons, or equivalently, 
the electron fraction, is crucial to the observed radiation 
from GRBs \citep{1999ApJ...521..640D,2002ApJ...573..770P}, its
dynamic evolution \citep{2000PhRvL..85.2673F,2003ApJ...588..931B} and 
the nucleosynthesis in the disk or fireball
\citep{2003ApJ...586.1254P,2003ApJ...588..931B}.
For instance, the inelastic collisions between 
neutrons and protons produce observable multi-GeV neutrino emission
\citep{2000PhRvL..85.1362B,2002ARA&A..40..137M}. In addition,
the two component fluid of neutrons and protons 
significantly changes the fireball interaction with an external medium
which is supposed to produce the observed electromagnetic radiation
from GRBs and their afterglows \citep{2003ApJ...585L..19B}; and
the electron fraction $Y_e$ strongly affects the equation of state of 
the hyperaccretion disk and the neutrino emissions from it \citep{2003sgrb}.

Roughly speaking, SNe are powered by the iron core collapse of their 
progenitors. Most numerical simulations have shown not only the failure of 
the prompt shock, but the failure of its revival by the delayed
neutrino emission from the protoneutron star (PNS). The result (explosion 
or not) sensitively depends on the input microphysics, such as the 
electron capture, the neutrino emission, neutrino-matter interactions,
and so on (see ref.\citep{2001pnsi.conf..333J}, and 
references therein). Without a doubt, 
weak interactions, especially $e^{\pm}$ capture and neutron 
decay, play a key role in both GRBs and SNe. During the accretion 
or collapse, these processes exhaust electrons, thus decrease the degenerate
pressure of electrons. Meanwhile, they produce neutrinos  which carry 
the binding energy away. 
Therefore, electron capture is crucial
to the formation of the bounce shock of SNe, and the resulting neutrino spectra
strongly influence the neutrino-matter interactions which are energy dependent
and are essential for collapsing simulations \citep{2003RvMP...75..819L}.

The existence of a hot state with nucleons is the 
the common characteristics of both GRBs and SNe, as well as
the PNS, the bounce shock and the early universe
\citep{2002RvMP...74.1015W}. In these systems,
the electron and positron captures 
are the two most important physical processes \citep{Pina64}. 
The steady state is achieved via 
the following beta reactions \citep{1967SvA....10..970I},
\bea
e^-+p &\rightarrow & n+ \nu_e, \label{eq:e_cap}\\
e^++n &\rightarrow & p+ \bar{\nu}_e, \label{eq:p_cap}\\
n     &\rightarrow & p+e^- + \bar{\nu}_e .  \label{eq:n_dec}
\eea
These beta reaction rates are calculated in 
the previous studies, usually under one of three approximations: 
the nondegenerate approximation \citep{Pina64}, 
the degenerate approximation 
(\eg \citep{1983bhwd.book.....S}), 
and the elastic approximation
in which there is no energy transfer to nucleons 
\citep{1985ApJS...58..771B}. 
In this letter, applying the structure function
formalism developed by \citet{1998PhRvD..58a3009R}(see also 
\citep{1998PhRvC..58..554B}), I derive the exact reaction 
rates of beta processes for all particles at arbitrary degeneracy.
In addition, I find an analytic expression for determining
the kinetic equilibrium between electron capture and positron
capture, which is efficient to determine the steady state of the hot 
matter with positrons. 

\paragraph*{Reaction rates.}
From Fermi's golden rule, the reaction rates of 
the processes (\ref{eq:e_cap})-(\ref{eq:n_dec}) read
\be
\lambda=2\int \prod_{i=1}^{4} \left[\frac{d^3\vec{p_i}}{(2\pi)^3} 
  \right](2\pi)^4
  \delta^{(4)}(P_i-P_f) |M|^2 \cal{F},  \label{eq:rate_g}
\ee
where $P_i=(E_i,\vec{p_i})$ denotes the four-momentum of
particle $i$ ($i=\nu_e/\bar{\nu}_e, e^-/e^+, n, p$), 
$p_i=|\vec{p_i}|$, and $P_i$ and 
$P_f$ are the total initial and final momentum, respectively.
$|M|^2(=G_{\rmscr{F}}^2\cos^2\theta_{\rmscr{C}}(1+3g_{\rmscr{A}}^2))$ 
is the transition rate averaged over the initial spins,
here $G_{\rmscr{F}} \simeq 1.436 \times 10^{-49}~\rmmat{erg}~ 
\rmmat{cm}^{3}$ is the Fermi weak interaction constant,
$\theta_{\rmscr{C}}$ $(\sin\theta_{\rmscr{C}}=0.231)$ 
is the Cabibbo angle, and 
$g_{\rmscr{A}}=1.26$ is the axial-vector coupling constant.
$\cal{F}$ denotes the final--states blocking factor.
For instance, in reaction (1),
$\cal{F}$=$f_e f_p (1-f_n)$, where $f_i$ is the Fermi-Dirac function of
particle $i$. 
In this letter, we assume that the emitted neutrinos can escape freely
from the system.
Using the structure function formalism developed by 
\citet{1998PhRvD..58a3009R}, 
the above integrations can be simplified into only three dimensional ones,
\bea
\lambda_{e^-p}&=&\frac{1}{8\pi^5}|M|^2m_nm_pT
	\int_0^{\infty}d E_{\nu} 
 	\int_{m_e-E_{\nu}}^{\infty} dq_0  \nonumber \\
	& \times &\int_{|p_e-p_{\nu}|}^{|p_e+p_{\nu}|}dq 
	E_{\nu} E_e f_eS_{p \rightarrow n}(q_0,q) ,
		\label{eq:rate_e_cap1} \\
\lambda_{e^+n}&=&\frac{1}{8\pi^5}|M|^2m_nm_pT
	\int_0^{\infty}d E_{\bar{\nu}} 
	\int_{m_e-E_{\bar{\nu}}}^{\infty} dq_0 \nonumber \\
	& \times &\int_{|p_{e^+}-p_{\bar{\nu}}|}^{|p_{e^+}+p_{\bar{\nu}}|}dq 
	E_{\bar{\nu}} E_{e^+} f_{e^+}S_{n \rightarrow p}(q_0,q) ,
		\label{eq:rate_p_cap1}\\
\lambda_{n}&=&\frac{1}{8\pi^5}|M|^2m_nm_pT
	\int_0^{\infty}d E_{\bar{\nu}} 
 	\int_{m_e+E_{\bar{\nu}}}^{\infty} dq_0   \nonumber \\
	& \times &\int_{|p_e-p_{\bar{\nu}}|}^{|p_e+p_{\bar{\nu}}|}dq 
	E_{\bar{\nu}} E_e (1-f_e)S_{n \rightarrow p}(-q_0,q) ,
		\label{eq:rate_n_dec1}
\eea
where $S_{i\rightarrow j}(q_0,q)$ is the so-called dynamic 
form factor or structure function
which characterizes the isospin response of the system 
\citep{1998PhRvD..58a3009R}. The expression
$S_{i\rightarrow j}(q_0,q)$ is given by
\be
S_{i\rightarrow j}(q_0,q)=\frac{z+\xi_-}{1-\exp(-z)}  \label{eq:form} , 
\ee
where
\bea
z&=&\frac{q_0+\mu_i-\mu_j}{T}, \\
\xi_-&=&\ln\left[\frac{1+\exp((E_-^i-\mu_i)/T)}{1+\exp(E_-^i+q_0-\mu_j)} 
\right], \\
E_-^i&=m_i+&\frac{m_j^2(q_0+m_i-m_j-q^2/2m_j)^2}{2m_iq^2},
\eea
where $\mu_i$ and  $m_i$ are the chemical potential and the mass of 
baryons, and $q_0$, $q$ 
denote the momentum and energy transfer. 
In Eqs.~(\ref{eq:rate_e_cap1})-(\ref{eq:rate_n_dec1}),
$E_e=q_0+E_{\nu}$,
$E_{e^+}=q_0+E_{\bar{\nu}}$, and 
$E_{e}=q_0-E_{\bar{\nu}}$, respectively. 
Equations~(\ref{eq:rate_e_cap1})-(\ref{eq:rate_n_dec1}) are valid for
nonrelativistic and noninteracting baryons \citep{1998PhRvD..58a3009R}.
Below the nuclear density, this is a good approximation.

Analogous to the analysis in \citet{1998PhRvD..58a3009R}, 
it is easy to obtain the previous results
in the nondegenerate and degenerate limits of baryons.
As an illustration, the electron capture rate in nondegenerate 
limit is shown below,
\be
\lambda_{e^-p}\simeq\frac{1}{2\pi^3}|M|^2 n_p\int_Q^{\infty}dE_e
	E_ep_e(E_e-Q)^2f_e, 
	\label{eq:rate_e_cap2} 
\ee
where $Q=m_n-m_p$ is the mass difference between neutron and proton,
$n_i=2(m_iT/2\pi)^{3/2}\exp({\eta_i})$ is the number density of neutrons
and protons in the nondegenerate limit, and $\eta_i=(\mu_i-m_i)/T$ is the
reduced chemical potential. 
The above approximate rates are frequently cited 
in the literature to discuss the 
kinetic equilibrium for $\beta$-processes and 
the emissivity of neutrino emission, 
even though its validity should be checked carefully
\cite{1967SvA....10..970I,2003ApJ...586.1254P,2003ApJ...588..931B}.

\begin{figure}
\includegraphics[width=2.7in]{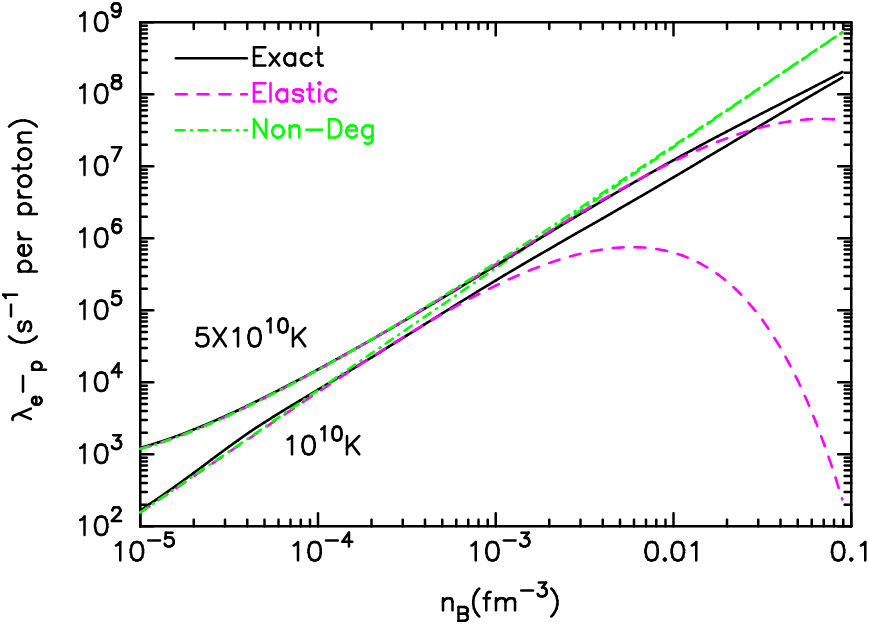}
\caption{Electron capture rates on protons in SNe matter as a function of
 the baryon number density $n_{\rmscr{B}}$ at different temperatures.
 The electron fraction $Y_e$ is taken to be 0.3. The solid curves 
 show the exact results from Eq.~(\ref{eq:rate_e_cap1}), the dashed 
 curves are the elastic results from Eq.~(\ref{eq:elastic}), 
 the dot-dashed curves are the nondegenerate results from 
 Eq.~(\ref{eq:rate_e_cap2}).
\label{fig:e_cap}
}
\end{figure}

In the elastic limit, 
$n_p$ in Eq.~(\ref{eq:rate_e_cap2}) is replaced by
\be
\eta_{pn}=(n_p-n_n)/(1-e^{(\eta_n-\eta_p)/T}). \label{eq:elastic}
\ee
It is generally believed that electron capture rate under 
the elastic approximation in some sense introduces the effects of the 
degeneracy of baryons, 
so it should be more accurate than that under the nondegenerate 
approximation. Suppose that the nuclei are dissolved completely 
into nucleons at high temperature. 
Figure~\ref{fig:e_cap} shows the differences between the
exact electron capture rate in the SNe matter
and the previous approximate results. The electron fraction 
$Y_e=(n_{e^-}-n_{e^+})/n_{\rmscr{B}}$ is assumed to be 0.3, 
where $n_{\rmscr{B}}=n_n(\mu_n,T)+n_p(\mu_p,T)$
is the baryon number density.
The number density of particles at any degeneracy is expressed
in terms of the Fermi-Dirac functions \citep{1998ApJS..117..627A}.
The multidimensional integrations and the Fermi-Dirac functions
are calculated using the mixture of Gauss-Legendre and Gauss-Laguerre
quadratures \citep{1998ApJS..117..627A}.
From Fig.~\ref{fig:e_cap}, it is evident that there are great differences between the elastic 
results and the exact results when baryons become degenerate.
As shown in Eq.~(\ref{eq:elastic}), 
the capture rates under the elastic approximation always decrease 
exponentially when nucleons become degenerate, which is 
qualitatively correct in the dense nuclear matter near 
$\beta$-equilibrium. 
However, this conclusion is not correct obviously in supernova matter. 
The elastic approximation is also good
when $\mu_p \simeq \mu_n$ because the favored energy 
transfer is zero. 
The elastic approximation underestimates the electron capture rate,
hopefully, the inclusion of the exact electron capture rate in the future 
collapsing simulations should be helpful for the supernova explosions.

\paragraph*{Condition for kinetic equilibrium.}
In principle, if the matter is transparent to neutrinos which carry
energy away, the equilibrium of the reactions 
(\ref{eq:e_cap})-(\ref{eq:n_dec}) can not be treated as a chemical 
equilibrium problem\citep{1983bhwd.book.....S}. 
Suppose that the dynamic time scale of the system under consideration is 
greater than that of the reactions
(\ref{eq:e_cap})-(\ref{eq:n_dec}), the general condition for the 
kinetic equilibrium is given by
\bea
&&\lambda_{e^-p}(\mu_n,\mu_p,\mu_e,T) \nonumber \\
&&=\lambda_{e^+n}(\mu_n,\mu_p,\mu_e,T)
+\lambda_{n}(\mu_n,\mu_p,\mu_e,T). \label{eq:equil}
\eea
However, 
it is well known that if all the particles involved in the Urca
processes are degenerate, such as what happens in the interior of a 
cold neutron star, then the typical energy of emitted neutrinos is of order 
the temperature, which could be neglected compared to the Fermi
energy of particles, \ie, $E_p+E_e=E_n$.
So the kinetic equilibrium requires $\lambda_{e^-p}=\lambda_{n}$
which results in the so-called chemical equilibrium condition
for the cold $npe^-$ gas,
\be
\mu_n=\mu_p+\mu_e. \label{eq:npe_eq}
\ee

Before we derive the analytic dynamical equilibrium condition 
in $e^{\pm}$ plasma with nucleons,
we can make some reasonable approximations. 
First, the rate of neutron decay
could be neglected before reaching the degenerate limit.
Second, the electrons are nondegenerate, 
\ie, $\exp((E_e-\mu_e)/T)\geq1$.
Third, the energy of emitted neutrinos is of order that of the captured
electrons/positrons, \ie, $E_{\nu}\simeq E_e$, $E_{\bar{\nu}}\simeq
E_{e^+}$, and thus $E_n\simeq E_p$. Under these approximations, 
$\lambda_{e^-p}=\lambda_{e^+n}$ gives
\bea
0&=&\lambda_{e^-p}-\lambda_{e^+n} \propto  f_p(1-f_n)f_{e^+} \nonumber \\
       && \times \left(\frac{e^{(E_e+\mu_e)/T}+1}
          {e^{(E_e-\mu_e)/T}+1} 
          -e^{((E_p-E_n)-(\mu_p-\mu_n))/T}\right) \nonumber \\
      &\simeq& f_p(1-f_n)f_{e^+}(e^{2\mu_e/T} -e^{(\mu_n-\mu_p)/T}) .
	\label{eq:app_nd}
\eea
Therefore, from Eq.~(\ref{eq:app_nd}) we obtain the beta-equilibrium 
condition for the $e^{\pm}$ plasma,
\be
\mu_n=\mu_p+2\mu_e . \label{eq:nd_eq}
\ee
It should be emphasized that during the above derivation, it is
not assumed whether the baryons are degenerate or not.
On the other hand, Eq.~(\ref{eq:nd_eq}) is still valid 
under the degeneracy of baryons, which is neglected 
completely in the approximate reaction rates 
at the beginning.
It is not a surprise to notice that 
the analytic equilibrium condition Eq.~(\ref{eq:nd_eq}) 
can also be drawn in the completely nondegenerate limit.
If all particles are nondegenerate, we have
\bea
\lambda_{e^-p} &\propto& 
           n_{e^-}n_p 
                \propto 
           \exp(\eta_p+\eta_e) 
	\label{eq:rate_e_cap3}, \\
\lambda_{e^+n} &\propto& 
           n_{e^+}n_n 
                \propto  
	 \exp(\eta_n-\eta_e) 
	\label{eq:rate_p_cap3} .
\eea
From Eq.~(\ref{eq:rate_e_cap3})-(\ref{eq:rate_p_cap3}), 
the analytic condition Eq.~(\ref{eq:nd_eq}) is also obtained
heuristically, but not very strictly.

If neutrinos are trapped, the chemical equilibrium condition 
for both cold $npe^-$ and hot $npe^{\pm}$ gases
is $\mu_n+\mu_{\nu_e}=\mu_p+\mu_e$. 
The chemical potential of the trapped neutrinos 
is generally assumed to be zero, thus, 
Eq.~(\ref{eq:npe_eq}) can also be understood as the chemical equilibrium
condition for $e^{\pm}$ plasma with neutrino trapping
\citep{2003ApJ...588..931B}.
As $T\rightarrow 0$, the number density of
trapped neutrinos $n_{\bar{\nu}_e}=n_{\nu_e}
\propto T^3 \rightarrow 0$.
Therefore, the difference between Eq.~(\ref{eq:npe_eq}) and 
Eq.~(\ref{eq:nd_eq}) clearly shows the effects of neutrino trapping.
\begin{figure}
\includegraphics[width=2.7in]{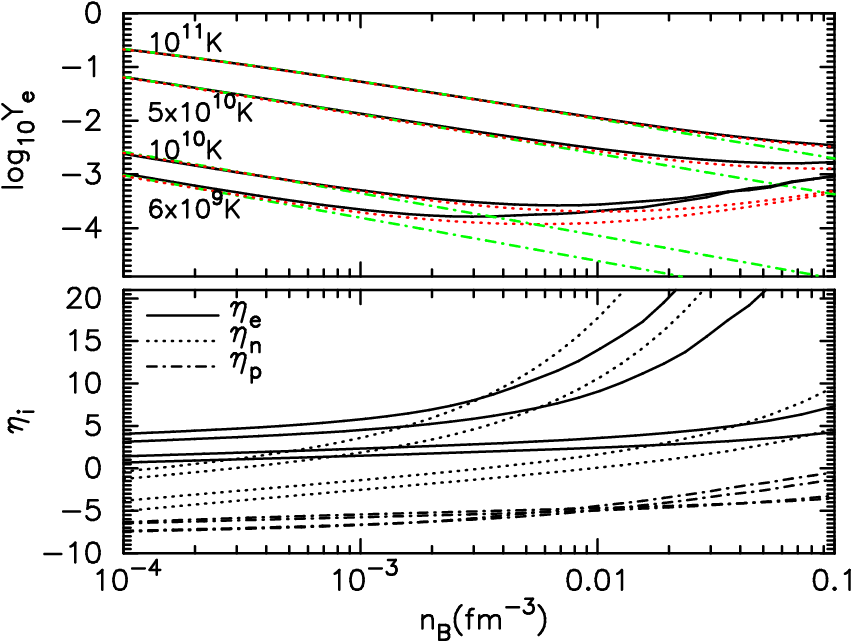}
\caption{
{\it Upper panel:} Electron fraction $Y_e$ under 
$\beta$-equilibrium
versus the baryon number density 
$n_{\rmscr{B}}$ at different temperatures. 
The solid curves show the exact results from 
Eqs.~(\ref{eq:rate_e_cap1})-(\ref{eq:rate_n_dec1}),
the dotted curves are from the analytic $\beta$-equilibrium 
condition Eq.~(\ref{eq:nd_eq}), and the dot-dashed curves are 
from the nondegenerate 
approximation of baryons.
{\it Lower panel:} The reduced chemical potentials of electrons, neutrons,
and protons $\eta_i$ as a function of the baryon number density at different 
temperatures. From top to bottom, the lines
correspond to the results of $T=6\times10^9$, $10^{10}$, $5\times10^{10}$, 
$10^{11}$K. \label{fig:frac}
}
\end{figure}
\paragraph*{Validity of the analytic condition.}
The  analytic condition Eq.~(\ref{eq:nd_eq}) is valid, only if 
electrons are nondegenerate, which is generally satisfied. 
If not, the number density of positrons will decrease exponentially.
In the following, we check Eq.~(\ref{eq:nd_eq}) by numerical calculations.
As before, suppose that the nuclei are dissolved completely 
into nucleons at high temperature. Given the baryon 
the baryon number density and temperature, the electron fraction 
$Y_e$ is determined by two equilibrium conditions. 
One is the charge neutrality 
$n_{e^-}(\mu_{e},T)-n_{e^+}(\mu_{e},T)=n_{p}(\mu_{p},T)$, 
the other is the beta equilibrium condition:
the integration equation 
Eq.~(\ref{eq:equil}), or our analytic equation Eq.~(\ref{eq:nd_eq}).
To check the 
validity of Eq.~(\ref{eq:nd_eq}), 
the equilibrium $Y_e$ is calculated based on the exact 
reaction rates Eqs.~(\ref{eq:rate_e_cap1})-(\ref{eq:rate_n_dec1}), 
the analytic condition Eq.~(\ref{eq:nd_eq}), and
the approximate rates,
respectively. 
Figure~\ref{fig:frac} shows $Y_e$ and the reduced chemical potentials $\eta_i$ 
versus the baryon number density at different temperatures. 
It is evident that the results from the analytic condition 
are almost consistent with the exact results. The difference occurs 
only in the 
regime where electrons become degenerate. 
As shown in the lower panel of Fig.~\ref{fig:frac},
neutrons become degenerate
as electrons do. The particles become degenerate if their reduced 
chemical potential exceeds the temperature. For electrons and 
neucleons, their degenerate
number densities are estimated to be
$
n_e^{\rmscr{deg}} \simeq T^3/3\pi^2
	\simeq9.0\times 10^{-6}T_{11}^3 
	\rmmat{fm}^{-3}$, 
and
$n_{n,p}^{\rmscr{deg}} \simeq (2m_{n,p}T)^{1.5}/3\pi^2
	\simeq2.8\times 10^{-3}T_{11}^{1.5} 
	\rmmat{fm}^{-3}$ 
respectively,
where $T_{11}=T/10^{11}$K.
In the same regime,
there are great differences (of several orders) between our results and
the the previous results
in which the degeneracy of nucleons is completely neglected.
In any case, Fig.~\ref{fig:frac} evidently shows that the results from the analytic condition 
are much more accurate than those from the approximate rates in all the 
parameter regions.

The advantage of having the analytic equilibrium condition at hand 
is obvious. For instance, we can derive some useful formulae under
the nondegenerate approximation. In such limit,
$n_n/n_p=1-1/Y_e=\exp((\mu_n-\mu_p-Q)/T)=\exp((2\mu_e-Q)/T)$,
Therefore, we have 
\be
n_{\rmscr{B}}Y_e=\frac{T^3}{3 \pi^2} \left[\left(\frac{\mu_e}{T}\right)^3+\pi^2
\left(\frac{\mu_e}{T}\right)\right]=\frac{n_{\rmscr{B}}}{1+e^{(2\mu_e-Q)/T}}.
\label{eq:app_mue}
\ee
The left equality is exact for the relativistic $e^{\pm}$
(\eg \cite{1996ApJS..106..171B}).
Based on Eq.~(\ref{eq:app_mue}), $\mu_e$ and $Y_e(\mu_e)$ can 
be determined. This simple result
completely reproduces the previous numerical results 
which is corresponding to the dot-dashed curves in Fig.~\ref{fig:frac}.
At $\mu_e/T<1$ and $(2\mu_e-Q)/T<1$,
Eq.~(\ref{eq:app_mue}) can be simplified further,
\be
Y_e=\frac{1}{2}\frac{(1+0.5Q/T)}{(1+1.5n_{\rmscr{B}}/T^3)}.
\label{eq:ye_app}
\ee
A similar result to Eq.~(\ref{eq:ye_app}) was obtained previously 
in a different method \citep{2003ApJ...588..931B}.

\begin{figure}
\includegraphics[width=2.7in]{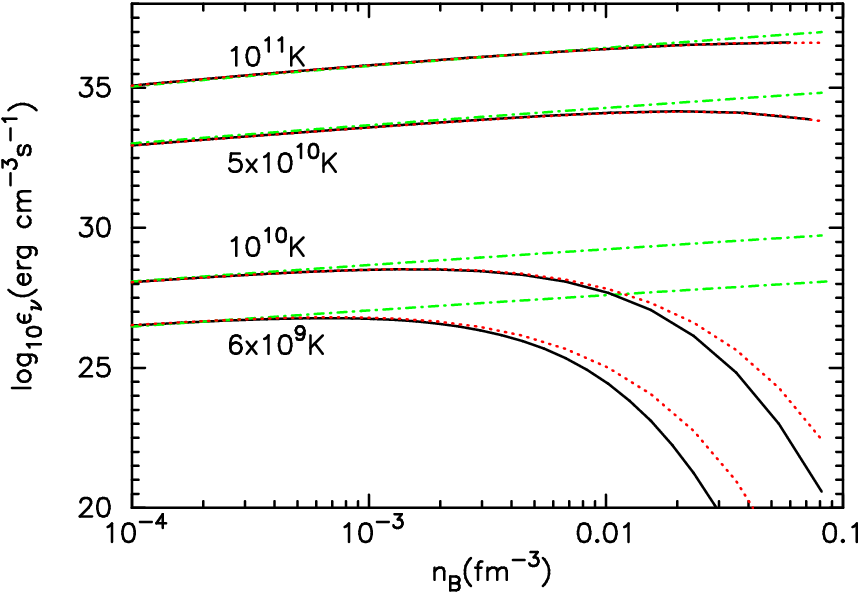}
\caption{
The neutrino emissivity $\epsilon_{\nu}$ versus the baryon number density at different 
temperatures. The line are the same as those of the upper 
panel in Fig.~\ref{fig:frac}
\label{fig:emis}
}
\end{figure}
The total neutrino emissivity under $\beta$-equilibrium
is shown in Fig.~\ref{fig:emis}.
Compared with the exact result, both approximate 
methods overestimate the rate of the neutrino emission because 
neglecting of the degeneracy of particles increases the phase space for 
the relevant reactions.
It is clearly shown in Fig.~\ref{fig:emis} that the effect of 
the neutron degeneracy 
is much more important than that of electrons, if such conditions are
satisfied.
 
\begin{acknowledgments}

The author would like to thank the anonymous referee 
for her/his constructive suggestions,
Dr. Ramesh Narayan, 
Dr. Jeremy Heyl and Dr. Rosalba Perna for many discussions,
Dr. Dong Lai for comments,
and Dr. David Rusin for a critical reading of this manuscript.
The author acknowledges the hospitality of Harvard-Smithsonian Center
for Astrophysics.  
This work is partially supported by the Special Funds for Major State
Research Projects, and the National Natural Science Foundation
(10233030).
\end{acknowledgments}


\begin{thebibliography}{21}
\expandafter\ifx\csname natexlab\endcsname\relax\def\natexlab#1{#1}\fi
\expandafter\ifx\csname bibnamefont\endcsname\relax
  \def\bibnamefont#1{#1}\fi
\expandafter\ifx\csname bibfnamefont\endcsname\relax
  \def\bibfnamefont#1{#1}\fi
\expandafter\ifx\csname citenamefont\endcsname\relax
  \def\citenamefont#1{#1}\fi
\expandafter\ifx\csname url\endcsname\relax
  \def\url#1{\texttt{#1}}\fi
\expandafter\ifx\csname urlprefix\endcsname\relax\def\urlprefix{URL }\fi
\providecommand{\bibinfo}[2]{#2}
\providecommand{\eprint}[2][]{\url{#2}}

\bibitem[{\citenamefont{{M{\' e}sz{\' a}ros}}(2002)}]{2002ARA&A..40..137M}
\bibinfo{author}{\bibfnamefont{P.}~\bibnamefont{{M{\' e}sz{\' a}ros}}},
  \bibinfo{journal}{\araa} \textbf{\bibinfo{volume}{40}}, \bibinfo{pages}{137}
  (\bibinfo{year}{2002}).

\bibitem[{\citenamefont{{Janka} et~al.}(2001)\citenamefont{{Janka},
  {Kifonidis}, and {Rampp}}}]{2001pnsi.conf..333J}
\bibinfo{author}{\bibfnamefont{H.}~\bibnamefont{{Janka}}},
  \bibinfo{author}{\bibfnamefont{K.}~\bibnamefont{{Kifonidis}}},
  \bibnamefont{and} \bibinfo{author}{\bibfnamefont{M.}~\bibnamefont{{Rampp}}},
  in \emph{\bibinfo{booktitle}{Physics of Neutron Star
  Interiors}} (\bibinfo{publisher}{Springer-Verlag}, 
  \bibinfo{address}{Heidelberg},\bibinfo{year}{2001}), pp. \bibinfo{pages}{363}.

\bibitem[{\citenamefont{{Narayan} et~al.}(1992)\citenamefont{{Narayan},
  {Paczynski}, and {Piran}}}]{1992ApJ...395L..83N}
\bibinfo{author}{\bibfnamefont{R.}~\bibnamefont{{Narayan}}},
  \bibinfo{author}{\bibfnamefont{B.}~\bibnamefont{{Paczynski}}},
  \bibnamefont{and} \bibinfo{author}{\bibfnamefont{T.}~\bibnamefont{{Piran}}},
  \bibinfo{journal}{\apjl} \textbf{\bibinfo{volume}{395}}, \bibinfo{pages}{L83}
  (\bibinfo{year}{1992});
\bibinfo{author}{\bibfnamefont{R.}~\bibnamefont{{Popham}}},
  \bibinfo{author}{\bibfnamefont{S.~E.} \bibnamefont{{Woosley}}},
  \bibnamefont{and} \bibinfo{author}{\bibfnamefont{C.}~\bibnamefont{{Fryer}}},
  \bibinfo{journal}{\apj} \textbf{\bibinfo{volume}{518}}, \bibinfo{pages}{356}
  (\bibinfo{year}{1999});
\bibinfo{author}{\bibfnamefont{R.}~\bibnamefont{{Narayan}}},
  \bibinfo{author}{\bibfnamefont{T.}~\bibnamefont{{Piran}}}, \bibnamefont{and}
  \bibinfo{author}{\bibfnamefont{P.}~\bibnamefont{{Kumar}}},
  \bibinfo{journal}{\apj} \textbf{\bibinfo{volume}{557}}, \bibinfo{pages}{949}
  (\bibinfo{year}{2001});
\bibinfo{author}{\bibfnamefont{T.}~\bibnamefont{{Di Matteo}}},
  \bibinfo{author}{\bibfnamefont{R.}~\bibnamefont{{Perna}}}, \bibnamefont{and}
  \bibinfo{author}{\bibfnamefont{R.}~\bibnamefont{{Narayan}}},
  \bibinfo{journal}{\apj} \textbf{\bibinfo{volume}{579}}, \bibinfo{pages}{706}
  (\bibinfo{year}{2002}).

\bibitem[{\citenamefont{{Pruet} et~al.}(2003)\citenamefont{{Pruet}, {Woosley},
  and {Hoffman}}}]{2003ApJ...586.1254P}
\bibinfo{author}{\bibfnamefont{J.}~\bibnamefont{{Pruet}}},
  \bibinfo{author}{\bibfnamefont{S.~E.} \bibnamefont{{Woosley}}},
  \bibnamefont{and} \bibinfo{author}{\bibfnamefont{R.~D.}
  \bibnamefont{{Hoffman}}}, \bibinfo{journal}{\apj}
  \textbf{\bibinfo{volume}{586}}, \bibinfo{pages}{1254} (\bibinfo{year}{2003}).

\bibitem[{\citenamefont{{Beloborodov}}(2003{\natexlab{a}})}]{2003ApJ...588..93%
1B}
\bibinfo{author}{\bibfnamefont{A.~M.} \bibnamefont{{Beloborodov}}},
  \bibinfo{journal}{\apj} \textbf{\bibinfo{volume}{588}}, \bibinfo{pages}{931}
  (\bibinfo{year}{2003}{\natexlab{a}}).

\bibitem[{\citenamefont{{Derishev} et~al.}(1999)\citenamefont{{Derishev},
  {Kocharovsky}, and {Kocharovsky}}}]{1999ApJ...521..640D}
\bibinfo{author}{\bibfnamefont{E.~V.} \bibnamefont{{Derishev}}},
  \bibinfo{author}{\bibfnamefont{V.~V.} \bibnamefont{{Kocharovsky}}},
  \bibnamefont{and} \bibinfo{author}{\bibfnamefont{V.~V.}
  \bibnamefont{{Kocharovsky}}}, \bibinfo{journal}{\apj}
  \textbf{\bibinfo{volume}{521}}, \bibinfo{pages}{640} (\bibinfo{year}{1999}).

\bibitem[{\citenamefont{{Pruet} and {Dalal}}(2002)}]{2002ApJ...573..770P}
\bibinfo{author}{\bibfnamefont{J.}~\bibnamefont{{Pruet}}} \bibnamefont{and}
  \bibinfo{author}{\bibfnamefont{N.}~\bibnamefont{{Dalal}}},
  \bibinfo{journal}{\apj} \textbf{\bibinfo{volume}{573}}, \bibinfo{pages}{770}
  (\bibinfo{year}{2002}).

\bibitem[{\citenamefont{{Fuller} et~al.}(2000)\citenamefont{{Fuller}, {Pruet},
  and {Abazajian}}}]{2000PhRvL..85.2673F}
\bibinfo{author}{\bibfnamefont{G.~M.} \bibnamefont{{Fuller}}},
  \bibinfo{author}{\bibfnamefont{J.}~\bibnamefont{{Pruet}}}, \bibnamefont{and}
  \bibinfo{author}{\bibfnamefont{K.}~\bibnamefont{{Abazajian}}},
  \bibinfo{journal}{Phys. Rev. Lett.} \textbf{\bibinfo{volume}{85}},
  \bibinfo{pages}{2673} (\bibinfo{year}{2000}).

\bibitem[{\citenamefont{{Bahcall} and {M{\' e}sz{\'
  a}ros}}(2000)}]{2000PhRvL..85.1362B}
\bibinfo{author}{\bibfnamefont{J.~N.} \bibnamefont{{Bahcall}}}
  \bibnamefont{and} \bibinfo{author}{\bibfnamefont{P.}~\bibnamefont{{M{\'
  e}sz{\' a}ros}}}, \bibinfo{journal}{Phys. Rev. Lett.}
  \textbf{\bibinfo{volume}{85}}, \bibinfo{pages}{1362} (\bibinfo{year}{2000}).

\bibitem[{\citenamefont{{Beloborodov}}(2003{\natexlab{b}})}]{2003ApJ...585L..1%
9B}
\bibinfo{author}{\bibfnamefont{A.~M.} \bibnamefont{{Beloborodov}}},
  \bibinfo{journal}{\apjl} \textbf{\bibinfo{volume}{585}}, \bibinfo{pages}{L19}
  (\bibinfo{year}{2003}{\natexlab{b}}).

\bibitem[{\citenamefont{{Yuan} et~al.}(2003)\citenamefont{{Yuan}, {Perna},
  {Di Matteo}, and {Narayan}}}]{2003sgrb}
\bibinfo{author}{\bibfnamefont{Y.}~\bibnamefont{{Yuan}}},
  \bibinfo{author}{\bibfnamefont{R.}~\bibnamefont{{Perna}}}, 
  \bibinfo{author}{\bibfnamefont{T.}~\bibnamefont{{Di Matteo}}},
  \bibnamefont{and}
  \bibinfo{author}{\bibfnamefont{R.}~\bibnamefont{{Narayan}}},
  \bibinfo{journal}{In preparation}  (\bibinfo{year}{2003}).

\bibitem[{\citenamefont{{Langanke} and
  {Mart{\'{\i}}nez-Pinedo}}(2003)}]{2003RvMP...75..819L}
\bibinfo{author}{\bibfnamefont{K.}~\bibnamefont{{Langanke}}} \bibnamefont{and}
  \bibinfo{author}{\bibfnamefont{G.}~\bibnamefont{{Mart{\'{\i}}nez-Pinedo}}},
  \bibinfo{journal}{Rev. Mod. Phys.} \textbf{\bibinfo{volume}{75}},
  \bibinfo{pages}{819} (\bibinfo{year}{2003}).

\bibitem[{\citenamefont{{Woosley} et~al.}(2002)\citenamefont{{Woosley},
  {Heger}, and {Weaver}}}]{2002RvMP...74.1015W}
\bibinfo{author}{\bibfnamefont{S.~E.} \bibnamefont{{Woosley}}},
  \bibinfo{author}{\bibfnamefont{A.}~\bibnamefont{{Heger}}}, \bibnamefont{and}
  \bibinfo{author}{\bibfnamefont{T.~A.} \bibnamefont{{Weaver}}},
  \bibinfo{journal}{Rev. Mod. Phys.} \textbf{\bibinfo{volume}{74}},
  \bibinfo{pages}{1015} (\bibinfo{year}{2002});
\bibinfo{author}{\bibfnamefont{M.}~\bibnamefont{{Prakash}}}
  \bibnamefont{\etal},
  \bibinfo{journal}{\physrep} \textbf{\bibinfo{volume}{280}},
  \bibinfo{pages}{1} (\bibinfo{year}{1997}).

\bibitem[{\citenamefont{Pinaev}(1964)}]{Pina64}
\bibinfo{author}{\bibfnamefont{V.}~\bibnamefont{Pinaev}},
  \bibinfo{journal}{Soviet Physics JETP} \textbf{\bibinfo{volume}{18}},
  \bibinfo{pages}{377} (\bibinfo{year}{1964});
\bibinfo{author}{\bibfnamefont{C.~J.} \bibnamefont{{Hansen}}},
  \bibinfo{journal}{\apss} \textbf{\bibinfo{volume}{1}}, \bibinfo{pages}{499}
  (\bibinfo{year}{1968}).

\bibitem[{\citenamefont{{Imshennik} et~al.}(1967)\citenamefont{{Imshennik},
  {Nadezhin}, and {Pinaev}}}]{1967SvA....10..970I}
\bibinfo{author}{\bibfnamefont{V.~S.} \bibnamefont{{Imshennik}}},
  \bibinfo{author}{\bibfnamefont{D.~K.} \bibnamefont{{Nadezhin}}},
  \bibnamefont{and} \bibinfo{author}{\bibfnamefont{V.~S.}
  \bibnamefont{{Pinaev}}}, \bibinfo{journal}{Soviet Astron.}
  \textbf{\bibinfo{volume}{10}}, \bibinfo{pages}{970} (\bibinfo{year}{1967}).

\bibitem[{\citenamefont{{Shapiro} and {Teukolsky}}(1983)}]{1983bhwd.book.....S}
\bibinfo{author}{\bibfnamefont{S.~L.} \bibnamefont{{Shapiro}}}
  \bibnamefont{and} \bibinfo{author}{\bibfnamefont{S.~A.}
  \bibnamefont{{Teukolsky}}}, \emph{\bibinfo{title}{{
  Black holes, white dwarfs, and neutron stars: 
  The physics of compact objects}}}
  (\bibinfo{publisher}{Wiley-Interscience}, \bibinfo{address}{New York},
  \bibinfo{year}{1983}).

\bibitem[{\citenamefont{{Bruenn}}(1985)}]{1985ApJS...58..771B}
\bibinfo{author}{\bibfnamefont{S.~W.} \bibnamefont{{Bruenn}}},
  \bibinfo{journal}{\apjs} \textbf{\bibinfo{volume}{58}}, \bibinfo{pages}{771}
  (\bibinfo{year}{1985}).

\bibitem[{\citenamefont{{Reddy} et~al.}(1998)\citenamefont{{Reddy}, {Prakash},
  and {Lattimer}}}]{1998PhRvD..58a3009R}
\bibinfo{author}{\bibfnamefont{S.}~\bibnamefont{{Reddy}}},
  \bibinfo{author}{\bibfnamefont{M.}~\bibnamefont{{Prakash}}},
  \bibnamefont{and} \bibinfo{author}{\bibfnamefont{J.~M.}
  \bibnamefont{{Lattimer}}}, \bibinfo{journal}{\prd}
  \textbf{\bibinfo{volume}{58}}, \bibinfo{pages}{13009} (\bibinfo{year}{1998}).

\bibitem[{\citenamefont{{Burrows} and {Sawyer}}(1998)}]{1998PhRvC..58..554B}
\bibinfo{author}{\bibfnamefont{A.}~\bibnamefont{{Burrows}}} \bibnamefont{and}
  \bibinfo{author}{\bibfnamefont{R.~F.} \bibnamefont{{Sawyer}}},
  \bibinfo{journal}{\prc} \textbf{\bibinfo{volume}{58}}, \bibinfo{pages}{554}
  (\bibinfo{year}{1998}).

\bibitem[{\citenamefont{{Aparicio}}(1998)}]{1998ApJS..117..627A}
\bibinfo{author}{\bibfnamefont{J.~M.} \bibnamefont{{Aparicio}}},
  \bibinfo{journal}{\apjs} \textbf{\bibinfo{volume}{117}}, \bibinfo{pages}{627}
  (\bibinfo{year}{1998}); see also http://flash.uchicago.edu/\~fxt/code\_pages
/fermi\_dirac.shtml.

\bibitem[{\citenamefont{{Blinnikov} et~al.}(1996)\citenamefont{{Blinnikov},
  {Dunina-Barkovskaya}, and {Nadyozhin}}}]{1996ApJS..106..171B}
\bibinfo{author}{\bibfnamefont{S.~I.} \bibnamefont{{Blinnikov}}},
  \bibinfo{author}{\bibfnamefont{N.~V.} \bibnamefont{{Dunina-Barkovskaya}}},
  \bibnamefont{and} \bibinfo{author}{\bibfnamefont{D.~K.}
  \bibnamefont{{Nadyozhin}}}, \bibinfo{journal}{\apjs}
  \textbf{\bibinfo{volume}{106}}, \bibinfo{pages}{171} (\bibinfo{year}{1996}).


\end{thebibliography}
\end{document}